# Prognostics and Health Management in Polymer Electrolyte Fuel Cells: Current Trends, Challenges, and Future Directions

Farideh Abdollahi [1,2], Kourosh Malek [1,3, *], Thomas Kadyk [1], Nadiia Kulyk [4], Christophe Gerling [4], Michael H. Eikerling [1,3,5]

[1] Theory and Computation of Energy Materials (IET-3), Forschungszentrum Jülich GmbH, 52425 Jülich/Germany
[2] RWTH Aachen University, Theory and Computation of Energy Materials, Faculty of Georesources and Materials Engineering, 52072 Aachen, Germany
[3] Centre for Advanced Simulation and Analytics (CASA), Simulation and Data Lab for Energy Materials, Forschungszentrum Jülich, 52425 Jülich, Germany
[4] EKPO Fuel Cell Technologies GmbH, Max-Eyth-Straße 2, D-72581 Dettingen and der Erms, Germany
[5] Chair of Theory and Computation of Energy Materials, Faculty of Georesources and Materials Engineering, RWTH Aachen University, 52062 Aachen, Germany

**\* E-mail:** k.malek@fz-juelich.de; ORCID: 0000-0002-3021-0813

**Abstract**

Prognostics and Health Management is crucial for the reliability and lifetime assessment of Polymer Electrolyte Fuel Cells (PEFCs). Here, we review the current advances on this topic, focusing mainly on key degradation mechanisms and methodologies such as physics-aware, data-driven, and hybrid modeling approaches. Key open challenges are analyzed, including the need for more accurate degradation modeling, effective management of multi-stack systems, and advancements in the currently underdeveloped action phase, in which diagnostic and prognostic insights are translated into real-time system responses, such as dynamic load derating, thermal-management adjustments, or automated maintenance triggers, to prevent failures and extend PEFC life. While notable strides have been made in recent years in diagnostics and remaining useful life estimation, it remains challenging to seamlessly integrate these insights into actionable strategies. Future directions highlight the need to address data scarcity and advance interdisciplinary research. Key focus areas include sensor integration, artificial intelligence, and digital twins. Additionally material innovations play a crucial role in bridging existing gaps. This work, therefore, intends to map the further development of Prognostics and Health Management systems toward ensuring the viability of PEFCs in practical applications.

**Keywords**

## 1. Introduction

Polymer Electrolyte Fuel Cells (PEFCs) continue to play a crucial role in future scenarios forging the development and commercial deployment of clean energy technologies due to their high efficiencies and substantial environmental benefits for applications in transport and stationary power systems [1]. Despite these advantages, commercialization of PEFCs has been limited in part because of the persistent issues of reliability and durability, which are especially sensitive under variable and dynamic conditions that these systems face under real-world operation. These challenges spur the development of prognostics and health management (PHM) systems that are important for achieving the required resilience of PEFCs. PHM entails the monitoring, prognosis, and management of the systems health to reduce the probability of unexpected failures, extend cell lifetimes, and optimize cell performance [2]. Cell "lifetime" itself is usually defined by an output-drop criterion: Pei et al. [3] take it as the time from initial operation until a 10 % decrease in rated-power voltage, while the U.S. Department of Energy likewise sets lifetime as the time for power to fall by more than 10 % of its initial rating [4]. Consequently, PHM is a valuable tool for PEFC technology development.

The PHM for PEFCs involves an integrated multistep process of continuous monitoring, diagnostics, and prognosis, which will interactively lead to optimum maintenance with appropriate operational adaptations in a timely manner. Figure 1 points out three major steps in the PHM process: observation, analysis, and action [5]. The observation stage involves a deep understanding of mechanisms of degradation of the device under test (DUT) by the researchers through intensive experimental studies and data gathering in a systematic manner. This forms the very basis for correctly determining the state of health (SOH) of the DUT. In the analysis phase, data collected must be critically examined to identify and diagnose the current SOH of the PEFC in order to predict its future behavior. PHM finally introduces the action that incorporates the insights from observation and analysis to drive a decision-making process, which ensures that efforts for maintenance or optimization are precisely targeted to the needs of the system. If the system operator needs to be notified or engaged, a Human-Machine Interface (HMI), such as an infotainment system, can act as the communication bridge between the system and the operator [6].

Early research in this domain, as represented by the works of Jouin et al. [2] and Hissel and Pera [7], laid the groundwork in terms of data acquisition and integration of sensors. Yet, these works had already initiated the exploration of diagnostic and prognostic techniques. Nowadays, the field has moved to advanced analysis methods and decision-making strategies, a fact that points toward a holistic approach of PEFC health management across the observation, analysis, and action phases. Recent research in the field of PHM of PEFC, as gleaned from published papers such as Wang et al. [8], demonstrates advanced analytical techniques for diagnostics and prognosis. Most are using machine learning (ML) and data-driven approaches to increase the accuracy of remaining useful life (RUL) prediction and explain the root cause of complex degradation behavior. Besides, there is a recent trend to develop hybrid approaches which will be able to bridge model-based and data-driven methodologies in handling nonlinear problems arising from PEFC degradations [9].

From an industrial standpoint, deployable AI models for PHM for PEFCs must satisfy practical requirements. First, they should generalize across wide operating ranges and dynamic conditions, and be robust to noise, missing data, and sensor faults, prerequisites for safety-critical fuel-cell systems [10]. Second, SOH/RUL solutions should be lightweight and fast enough for onboard integration in automotive and stationary applications; some accuracy–latency trade-offs are acceptable, provided models remain reliable under variable loads, start-up/shut-down cycles, and environmental changes [11]. Third, interpretability is crucial: models should not only estimate remaining life of a device but also localize degrading components and provide recommendations on diagnostics and maintenance. Specifically for fuel cell stacks, differentiation between reversible and irreversible degradation of components and recommendations on recovery measures that can be done onboard are of interest [10]. Finally, aside from the purely degradation-related aspect of AI models for PEFC, the possibility to implement virtual sensors also opens the door to better online monitoring of important parameters and optimization of system control strategy to enhance system performance/efficiency [12]. For instance, a physical model of a stack can be used to compute virtual sensors which are not measured in the real system and these quantities can then be integrated into an ML model to enrich real-life system data, since physical models often require substantial computational resources and are not suitable for real-time applications. Specifically, a high-fidelity stack model can be used offline to generate unmeasured quantities (virtual-sensor targets) and enrich the training data; a trained, lightweight ML estimator then runs online on measured signals to infer these quantities in real time, avoiding the need to solve the full physical model during operation.

Current literature shows a high emphasis on RUL estimation, especially in the prognostics stage of the PHM analysis phase. While diagnostics and prognostics have immensely enhanced the improvement of the analysis phase, less literature has fully explored the action phase. This phase is important for the support of real-time adaptive responses to dynamic operational conditions. Secondly, most of the works that exist still have scant comprehensive frameworks on integrating such insights into effective decision-making processes.

There is currently a notable gap in existing methodologies for feature extraction and condition assessment in PEFCs. Specifically, present monitoring approaches predominantly rely on signals such as cell voltage and stack power. While these system-level signals are effective for broad operational assessments, they are inadequate for precise RUL predictions at the component level. The gap lies in the insufficient attention and limited monitoring capabilities specifically targeted at critical PEFC components, including the Polymer Electrolyte (PEM) and the Cathode Catalyst Layer (CCL) within the Membrane Electrode Assembly (MEA). Complementarily, virtual sensors and physics-informed surrogates can supply component-level observables (e.g., local membrane hydration or CCL oxygen-transport metrics) to strengthen condition assessment where direct instrumentation is impractical [12].

Thus, current diagnostic and prognostic practices fail to accurately capture the degradation mechanisms and state of health specific to these individual components. Bridging this gap requires developing and employing advanced condition monitoring techniques that directly monitor critical component-specific indicators and degradation patterns. Such a holistic and integrated PHM framework would transcend the limitations of isolated diagnostics and prognostics, supporting continuous, real-time decision-making based directly on detailed component health insights. This implies coupling physics-aware, interpretable, and robust AI with decision logic capable of recommending component-targeted maintenance or recovery actions when appropriate.

Accordingly, we emphasize industrially relevant requirements, generalization, robustness to data issues, computational efficiency for onboard use, interpretability, and virtual sensing, to align PHM outcomes with the needs of real-world deployment. In this context, this work underlines key degradation mechanisms during operation of PEFCs under real world conditions and some of the challenges that exist in the action phase, system-level diagnostics and prognostics, and data acquisition. Hence, attention to the development of PHM methodologies is considered a cardinal area to be developed to support the deployment of PEFCs into real-world high-demand applications. By outlining these gaps, this review summarizes recent developments in this field and provides a roadmap for future research in developing robust dynamic PHM solutions for PEFC systems.

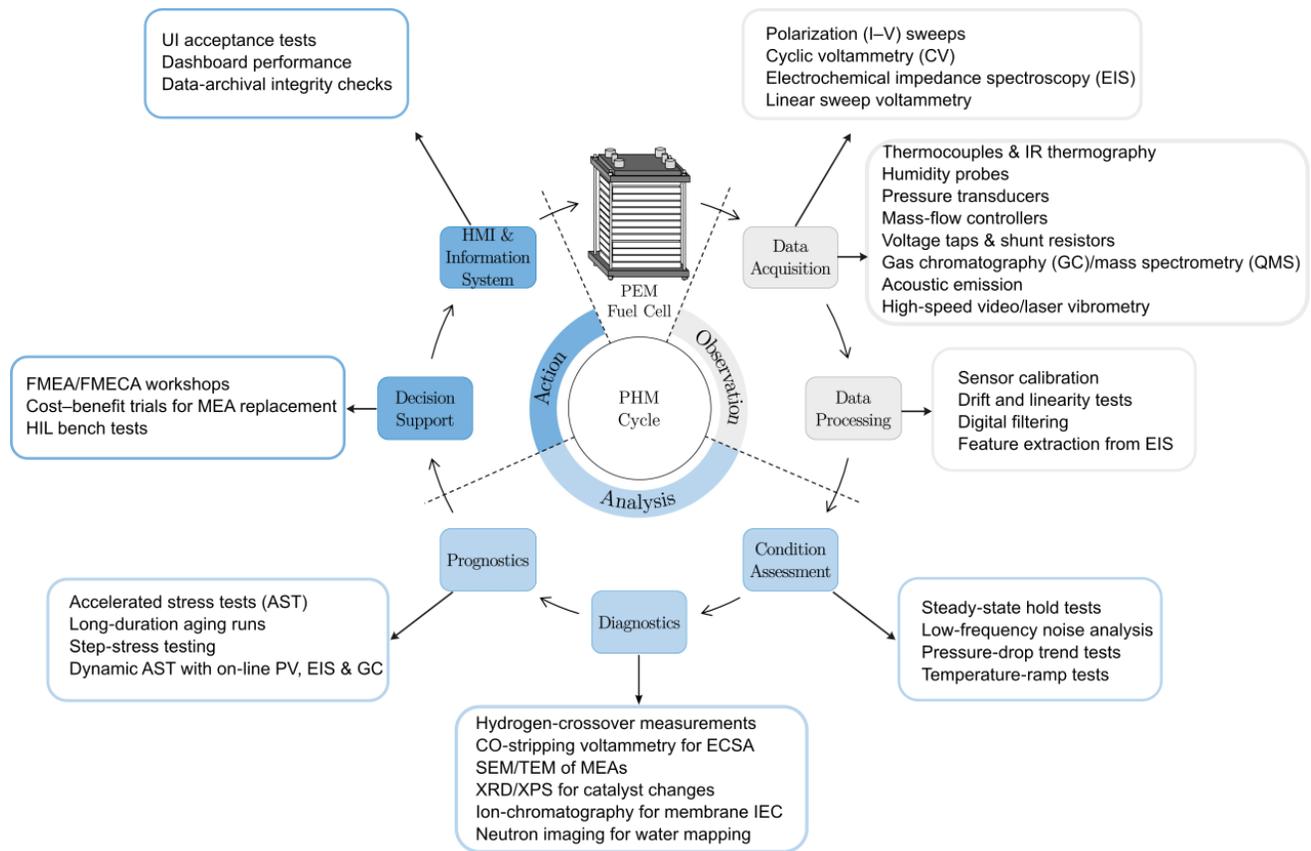

Figure 1. Schematic of the PHM cycle showing seven phases in three different categories [2, 6]. The Data Acquisition phase [13-21] gathers raw electrochemical and sensor data; Data Processing [22-25] converts and filters signals; Condition Assessment [26-28] evaluates current health; Diagnostics [29-36] pinpoints root causes; Prognostics [37-40] projects future degradation; Decision Support [41-45] translates insights into maintenance actions; and HMI & IS [46-49] delivers information to operators.

## 2. Degradation Pathways of PEFCs During the Observation Stage

The observation stage of PHM focuses on identifying and analyzing degradation mechanisms by experimentally monitoring critical signals and operational behaviors, providing a foundation for understanding how specific conditions contribute to the accelerated wear of PEFC components. PEFCs for automotive applications are subjected to some unique challenges such as dynamic operating conditions, variable loads, frequent start-stop cycles, idling, high current demands, and harsh environmental conditions, all of which accelerate the aging of the critical components such as PEM, CL, and GDL, which have direct influence on the performance and durability of fuel cells [50, 51]. Understanding of mechanisms degradation and their main drivers will help to enhance the durability and reliability of the PEFCs necessary for sustainable use in automotive applications. This section presents a focused review of key degradation mechanisms relevant to PEFCs. Comprehensive reviews on PEFC degradation have been covered elsewhere [52-54]. This section selectively addresses specific degradation aspects pertinent to PHM.

### 2.1. Membrane Degradation

For automotive applications, the PEM constitutes a critical component in PEFC that separates half-reactions, conducts protons, and prevents short-circuiting by blocking electron and reactant transport [55]. The hybrid-type design of PEMs, such as Nafion®-based membranes integrated with expanded microporous PTFE (ePTFE), has facilitated high-power densities and durability demanded by automotive fuel cells, as demonstrated, for instance in the Toyota Mirai [56, 57]. However, the harsh and dynamic operating conditions in automotive applications pose significant challenges to the long-term performance of these membranes. Prolonged exposure to extreme environments leads to inevitable degradation, driven by both chemical and mechanical factors. In this section, these two main degradation mechanisms in PEMs will be explained.

#### 2.1.1. Chemical Degradation

The chemical degradation of PEMs has been related to an attack by highly reactive radicals such as hydroxyl (HO$^\bullet$) and hydroperoxyl (HOO$^\bullet$) that compromise the structure of the polymer [58, 59]. Active radicals are mainly generated when hydrogen peroxide reacts with trace metal ions such as $Fe^{2+}$ and $Cu^{2+}$, which are generally introduced through the corrosion of bipolar plates or via contaminants in air and water streams [60-62]. These radicals attack the backbone and side chains of the ionomer membrane, leading to thinning, cracking, lamination, and pinhole formation that accelerate gas crossover and may lead to catastrophic failure of the MEA. Research works show that degradation is accelerated under both open circuit-voltage (OCV), and under low-humidity conditions, because in each case gas crossover increases [52]. The damaging effect of OCV-like conditions is usually quantified by measuring the fluoride emission rate and higher average HO$^\bullet$ density at low load current [63, 64]. Historically, research has proved that platinum (Pt) migration, which is most pronounced under open-circuit/idling conditions due to extensive Pt dissolution, leads to the formation of a Pt band in the membrane as well as reduced cathode Pt content and electrochemical surface area (ECSA) [63, 65, 66]. This obstructs the oxygen reduction reaction (ORR), impairs proton conductivity, and may cause internal short circuits, ultimately contributing to membrane degradation [67]. Additionally, as $Pt^{2+}$ migrates through the membrane's ion channels and precipitates as crystallites, it can obstruct proton transport and diminish the overall proton conductivity. However, the Pt band's role in membrane durability is debated. Some studies suggest that it accelerates degradation by promoting radical formation via $H_2O_2$ (through crossover reactions between oxygen and hydrogen) [68-70], while others propose that it mitigates damage by scavenging radicals and catalyzing water formation [71, 72].

Recent advances in PEM design have significantly reshaped the degradation landscape of PEFCs [54]. The use of reinforced membranes, such as those incorporating expanded PTFE or fiber mats, with reduced thicknesses (as low as 10 μm) and enhanced mechanical stability, combined with the incorporation of radical scavengers [73, 74], has improved their chemical durability and reduced the susceptibility to radical-induced damage. These innovations also appear to mitigate platinum (Pt) band formation by limiting Pt ion mobility and suppressing local radical concentrations.

For instance, Sgarbi et al. [75] investigated MEAs assembled with a 15 μm reinforced PFSA membrane and subjected them to a multiple-stressor accelerated stress test (AST) using a segmented PEFC. Across a range of cathode Pt loadings, they observed uniform $Pt^{2+}$ ion migration through the membrane, followed by consistent re-deposition near the cathode-membrane interface due to hydrogen crossover. Crucially, neither Pt band formation within the membrane bulk nor significant carbon corrosion was detected. These findings suggest that, under modern membrane and operating conditions, classic degradation effects such as deep Pt band formation may be largely suppressed. This highlights the impact of contemporary membrane architectures in overcoming some of the key durability challenges seen in earlier PEFC generations. While it remains useful to understand the older mechanisms associated with membrane failure, current evidence suggests that with state-of-the-art membranes, Pt band formation plays no role in degradation. Ongoing research is focusing instead on optimizing radical scavenger efficiency, membrane-electrode interactions, and long-term durability under dynamic operational conditions.

### 2.1.2. Mechanical Degradation

Thermal and humidity cycling under dynamic load conditions intensifies mechanical degradation in PEMs. Fluctuating operating conditions, including rapid transitions between low and high current densities, cause repetitive swelling and shrinking due to continuous hydration-dehydration cycles. These cyclic stresses induce mechanical fatigue, leading to cracks, pinholes, and membrane thinning, which subsequently increase gas crossover and reduce performance [76-78]. Furthermore, variation in  water content induced by these dynamic conditions significantly impacts proton conductivity [79]. Prolonged exposure to humidity cycling has been shown to cause permanent physical damage, as evidenced by increased hydrogen crossover current densities and visible cracks [80, 81]. Figure 2 schematically illustrates how different operating conditions affect membrane degradation.

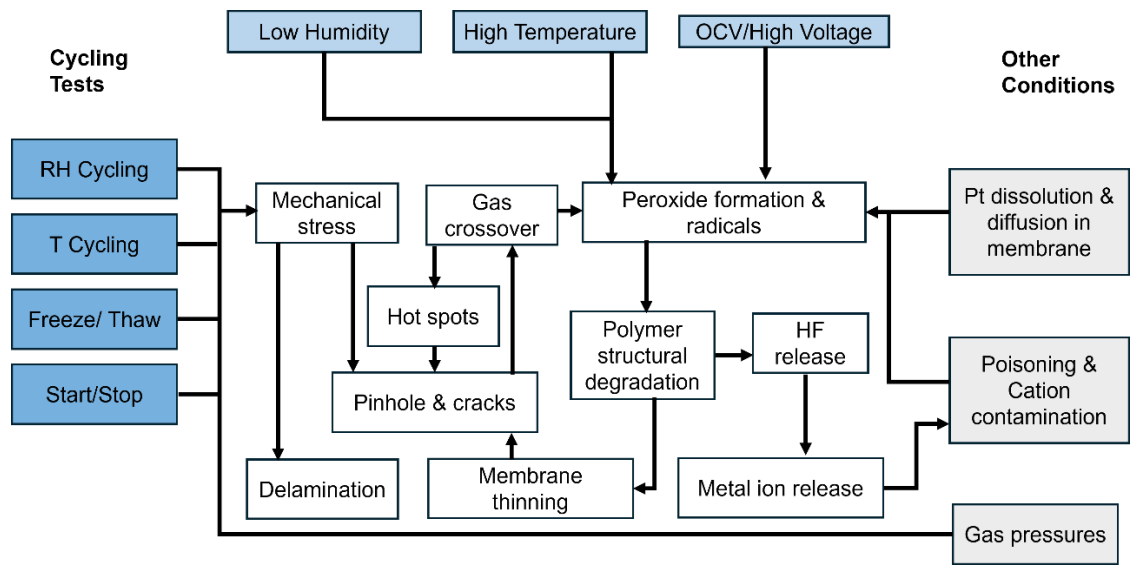

Figure 2. Membrane degradation under different accelerated stress test (AST) conditions [82].

## 2.2.  Catalyst Layer (CL)

Catalyst layers are other important components in PEFCs, which directly determine the efficiency of hydrogen oxidation reaction (HOR) and ORR and the performance of the fuel cell. Under dynamic operating conditions, such as start-stop cycles, load variations, and changes in OCV states, their durability faces tremendous challenges due to the accelerated degradation caused by a complex interplay of mechanical, thermal, chemical, and electrochemical drivers. Understanding this interplay is essential in the quest for enhancing the durability and reliability of PEFCs. In this section, we focus on the Pt catalyst in the cathode, which is particularly susceptible to degradation and faces the most significant challenges.

The degradation of the CCL under OCV conditions is mainly attributed to high electrode potentials, which favor the dissolution, migration, and re-accumulation of Pt, e.g., in the form of a Pt band in the PEM.  At cathode potential above 0.9 V, Pt nanoparticles dissolve into $Pt^{2+}$ ions [83, 84]. These ions can diffuse through the ionomer phase into the membrane, where they may precipitate and form a concentrated Pt band. This phenomenon leads to a loss in the electrochemically active surface area (ECSA) and it diminishes overall catalytic efficiency by reducing the availability of Pt for the ORR [85-87].   However, as highlighted in Section 2.1.1, recent advancements in membrane technology, such as the adoption of thin, reinforced membranes and the inclusion of radical scavengers, have been shown to significantly limit Pt ion mobility and suppress Pt band formation. These developments suggest that degradation caused by the Pt band, while historically important, may be substantially mitigated in modern PEFC systems.

High cathode potentials not only accelerate Pt dissolution but also enhance particle growth through Ostwald ripening and agglomeration since Pt redeposits on larger particles to lower the surface excess energy [87, 88]. Dynamic load and voltage cycling are two other important automobile conditions that impact both CCL degradation and ECSA loss. ECSA loss is further related to the load cycling that acts on a catalyst layer and  induces ionomer redistribution as well as the crack formation under mechanical stresses [89]. These cracks cause Pt particles near the damaged regions to detach from their carbon support and be carried away by water flow, reducing the active catalytic surface [90]. Simultaneously, carbon corrosion during periods of gas starvation weakens the structural integrity of the catalyst layer, and it further promotes the detachment and clustering of Pt particles [91]. Potential cycling further accelerate these mechanisms by enhancing the dissolution and redeposition of Pt particles through electrochemical Ostwald ripening [92, 93]. This results in nonuniform particle growth, a decrease in the number of active sites, and an increase in kinetic losses. Cycling between high and low potentials boosts these effects since high potentials promote Pt oxide formation and subsequent transition to low potentials and oxide reduction triggered by it enhance dissolution [94], while lower potentials favor particle agglomeration [95, 96]. A schematic illustrating the effect of different operating conditions on degradation mechanisms in the CL is shown in Figure 3.

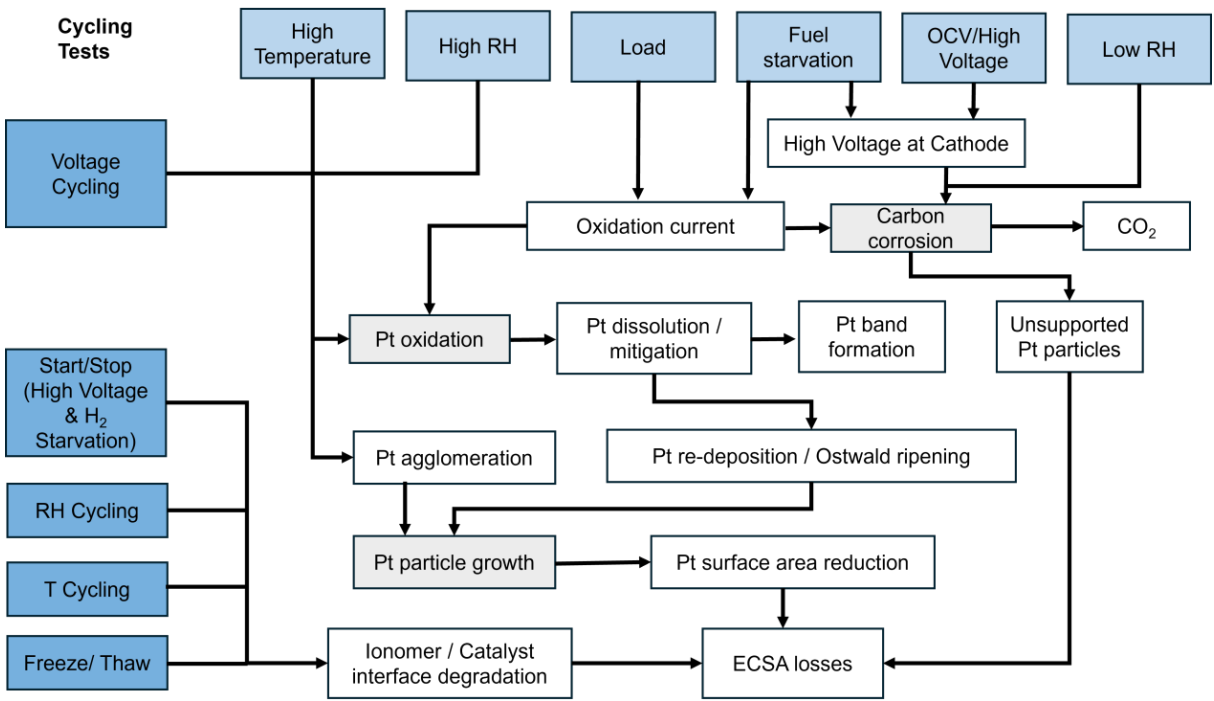

Figure 3. Catalyst degradation under different AST conditions [82].

## 2.3. Gas Diffusion Layer (GDL)

The GDL plays a very important role in maintaining effective mass, electron and heat transport, mechanical support, and water management in PEFCs, while it is particularly vulnerable to deterioration processes such as carbon corrosion and PTFE loss under different driving conditions [97-99]. Dynamic load cycles induce mechanical wear due to cyclic stresses and humidity fluctuations, which may lead to cracks, structural changes, and a loss of hydrophobicity, delamination of CL and GDL and thus to a deterioration of PEFC performance [100]. Startup and shutdown cycles further contribute to degradation via localized stresses caused by rapid temperature and potential transitions, resulting in carbon corrosion and mechanical failure [101]. High power or overload conditions create thermal hotspots and water management issues, which accelerate the loss of hydrophobicity and the risk of flooding [102]. These combined degradation mechanisms underline the need to design robust GDL materials and optimize operational strategies for improved PEFC durability and reliability under dynamic automotive conditions.

## 3. Analysis Stage

The analysis phase of PHM plays a pivotal role in transforming observational data into actionable insights for performance optimization and degradation mitigation in PEFCs. This phase focuses on identifying patterns, quantifying degradation rates, and correlating operational conditions with failure mechanisms through advanced analytical techniques. This section introduces the key states of health indicators and classifies prognostics methods.

### 3.1. Health Indicators (HIs)

In the field of fuel cell health monitoring, various Health Indicators (HIs) have been developed for diagnosing the degradation of PEFCs under steady-state and dynamic operation. Conventional HIs, such as voltage, current, and OCV, are widely employed due to their simple and well-defined measurement process and the effectiveness in characterizing the degradation state at the cell/stack level. ECSA is another HI which is typically assessed to capture catalyst activity and loss [103-109]. For instance, during constant current operation, it is commonly observed that the voltage at the fixed current density progressively declines. Although episodes of partial performance recovery may occur, influenced by specific operational conditions, these temporary improvements typically do not reverse the underlying degradation mechanisms. Numerous studies have revealed correlations between the deterioration of gradual reduction in these performance indicators and the aging process of PEFCs. [110].

### 3.1.1. Voltage Indicators

Voltage loss serves as a reliable indicator of fuel cell health. Jouin et al. [110] developed a particle filtering framework that models voltage loss as a degradation indicator for estimating the RUL of PEFCs. Ma et al. [111] applied data-driven methods such as t-distributed stochastic neighbor embedding and extreme gradient boosting to extract feature information from voltage data, enabling precise diagnostics of degradation types within a single cell. Similarly, Benouioua et al. [112] explored voltage singularity spectra, employing k-nearest neighbors to detect degradation through extracted singularity features for a PEFC Stack. Instead of voltage, indicators of fuel cell health could consider power signals as well. Power signal analysis includes the use of sparse autoencoders and deep neural networks, as demonstrated by Liu et al. [113] in their RUL prediction models. Ibrahim et al. [114] combined discrete wavelet transforms with power data, facilitating life time predictions in both static and dynamic conditions.

### 3.1.2. Dynamic Condition Monitoring

Dynamic operating conditions introduce fluctuating parameters, making it challenging to identify suitable HIs through conventional methods. To address this difficulty, methodologies such as virtual steady-state voltage, proposed by Li et al. [115] [14, 136], enable real-time tracking of voltage behavior under varying loads. Furthermore, advanced signal processing techniques have been employed to enhance predictive capabilities of these HIs under dynamic conditions. Methods such as the Hilbert-Huang transform [8] and empirical mode decomposition [116] have been used to extract degradation features from voltage signals, allowing for the detection of subtle changes that may indicate performance decline. Mousa et al. [117] applied Electrochemical Impedance Spectroscopy (EIS) to identify voltage drops associated with operational failures. Additionally, Hua et al. [118] proposed the Relative Power-Loss Rate (RPLR) for dynamic monitoring, which effectively guides RUL prediction by tracking deviations from initial power levels over time. Zhang and Pisu consider ECSA as a HI in their study to monitor the performance of PEFCs [104]. Cyclic voltammetry is used to measure the ECSA. However, it is primarily a laboratory-based characterization technique and cannot be implemented online, which limits its practical utility in real-time fuel cell monitoring.

### 3.1.3. Hybrid Health Indicators

Hybrid HIs, which integrate multiple indicators, provide a more comprehensive and reliable view of degradation by capturing complex correlations between parameters. Liu et al. [119] developed hybrid HIs combining exchange current, electronic resistance, and ionic resistance to monitor fuel cell aging with improved accuracy. Additionally, Chen et al. [120] introduced a fusion of stack voltage, internal resistance, and power indicators in a hybrid HI using a second-order Gaussian degradation model, achieving RUL prediction through unscented particle filtering. Hua et al. [121] utilized an Echo State Network for multi-input predictions, incorporating parameters such as stack voltage, current, temperature, and gas pressure to predict RUL. Innovative methodologies continue to emerge for deriving aging parameters within the PEFC degradation model. Lv et al. [122] proposed an HI that integrates the degradation model over specified current density ranges to encapsulate the effects of aging comprehensively. Wang et al. [123] combined EIS and polarization curves with Mahalanobis Distance for accurate health assessment. The primary challenge with hybrid methods lies in their computational complexity and the difficulty of determining appropriate weights for each individual indicator, a task that requires careful optimization and is far from being straightforward.

In conclusion, while steady state HIs remain valuable for diagnosing PEFC degradation under controlled conditions, dynamic and hybrid HIs have introduced new possibilities for monitoring performance under real-world fluctuating conditions. However, it is essential to prioritize methods that minimize complexity, do not disrupt the operation of the cell or stack, and accurately reflect the underlying degradation processes. Continued research in this direction will enhance our ability to reliably predict degradation and extend the operational life of PEFC systems in practical applications, particularly in automotive settings.

## 3.2. Prognostic Approaches for PEFCs

Existing studies on fuel cell prognostics can be classified into three types: model-based, data-driven, and hybrid prognostics.

### 3.2.1. Model-Based Approaches

Modeling fuel cell performance draws on a wide range of approaches across different scales and methodologies, from mechanistic models based on physical principles to empirical and semi-empirical approaches that use experimental data to improve predictability [124]. Equivalent circuit models are an approach that represents the impedance and voltage response of the fuel cell by using electrical analogs of resistors and capacitors, thus enabling speedy diagnostics of the electrical behavior without physical or chemical modeling [125]. Mechanistic models vary from atomic-scale models, to component-level models focusing on membrane durability, catalyst or GDL degradation mechanisms and system-level models addressing the operational dynamics of the whole fuel cell stack under a variety of conditions [126-128]. This review narrows its focus to models explicitly used in PHM, discussing the state-of-the-art tools present for monitoring and managing fuel cell health. However, a large number of other existing models at different scales, developed for performance prediction, may turn out to be very promising in view of integration within future health management frameworks. The adaptation of such multiscale models improves the predictive accuracy substantially and, hence, provides more detailed insight into both the performance and degradation dynamics in PEFC.

Mechanistic models are invaluable in PHM because of their precision, minimum training data requirement, and generalizability [129]. In addition to information on RUL, the user can concurrently view changes in the internal state and important parameters of PEFCs. These models incorporate relevant physical, chemical, and electrochemical phenomena, such as materials properties and degradation mechanisms, which can be simulated with a reasonable degree of accuracy [130]. The first published paper to estimate the RUL of PEFCs was presented by Zhang et al. [131], who proposed a component-level model focusing on catalyst degradation. This groundbreaking study established a physical model that links operating conditions to the reduction in ECSA as a primary degradation indicator. To achieve real-time tracking and RUL prediction under automotive load cycles, the authors implemented an Unscented Kalman Filter (UKF), setting a significant precedent in the field. Yet, the prognostic assessment was performed on a single cell rather than the entire stack and it covered a relatively short period of time (300 hours) [131].

In another study, Mayur et al. applied a multi-scale mechanistic model to predict the lifetime of a PEFC stack under automotive load cycles. This model targeted Pt degradation and tracked the loss of ECSA as a key indicator of Pt degradation via dissolution [132]. In another study, Bernhard et al. employed a two-part model to analyze the degradation in PEFC cathodes at the MEA level. This model consists of an ECSA-based statistical physics submodel and a Tafel kinetics-based performance submodel. By tracking ECSA, Tafel slope, and exchange current density, the model predicts voltage losses and catalyst degradation under varied AST conditions [133]. Dhanushkodi et al. developed a simplified kinetic model to predict carbon corrosion and performance loss in PEFCs. By applying ASTs with square-wave voltage cycles at different upper limits, the authors measured carbon loss and performance degradation in the CCL. Carbon loss was assessed through $CO_2$ evolution, while performance loss was evaluated based on voltage reduction at a set current density [134]. Polverino and Pianese also describe a model-based PHM framework intended for component level and to monitor degradation by analyzing the ECSA losses due to primary degradation mechanisms such as platinum dissolution and Ostwald ripening [135]. Ao et al. introduced a novel lifetime prediction method for PEFCs at the component or stack level, specifically under real-world power output cycles, by utilizing two distinct ECSA degradation models. The approach combined a physics-based, steady-power model that quantifies Pt dissolution at constant voltage, and an empirical, transient-power model that captures the extra ECSA loss induced by voltage swings via data fitting [109]. Together, these studies demonstrate the versatility and accuracy of mechanistic models in addressing various PEFC degradation phenomena under realistic operating conditions, making them essential tools in PHM. However, the complexity of degradation processes requires a detailed grasp of PEFC degradation mechanism at play, which poses a challenge. This complexity, along with significant computational demands of detailed mechanistic models, limits the widespread adoption of online prognostics. Furthermore, some PEFC degradation mechanisms remain poorly understood, and model parameters associated with these mechanisms can only be established by means of experimental data or from the expert judgment of PEFC specialists [128, 131, 133, 135].

Empirical models, derived solely from experimental data, are commonly employed for PEFC prognosis due to their simplicity and ease of implementation [136]. Unlike physics-based models, empirical approaches do not account for the microstructural complexities of the system. Instead, they rely on observer-based prognostic techniques to predict future behavior, such as the Kalman filter [137-139], or particle filter [110, 140]. Pei et al. developed and validated multiple lifetime prediction formulas for fuel cells, designed to support both laboratory-based and real-time, vehicle-based applications. The study introduced a maximum service lifetime formula and a nonlinear lifetime prediction formula and verified it against experimental data from single-cell and stack configurations as well as real-world vehicle operating conditions [141]. Although empirical models offer benefits a high of computational efficiency and ease of implementation, they often lack the precision and physical relevance needed to accurately reflect fuel cell degradation [142].

Semi-empirical models merge the strengths of both mechanistic and empirical approaches by using experimental data as a foundation while determining certain parameters through physical laws. These models typically maintain physical relevance through the HIs they employ [143]. Kneer et al. employed a semi-empirical catalyst degradation model to study the impact of automotive operating conditions on catalyst degradation at the MEA scale [144]. Ou et al. proposed a voltage prediction model that can be applied to degradation prediction and RUL estimation for the PEFC systems working under vehicle operating conditions based on a semi-empirical model [145]. A complete list of both empirical and semi- empirical models has been reviewed in Hua et al. [9].

### 3.2.2. Data-Driven Approaches

As fuel cell technology evolves, the need for more efficient and reliable prognostics methods is becoming increasingly important. The increasing reliance on data-driven methods in prognostics PEFC is mainly because these methods are flexible. For example, data-driven models bypass prior knowledge of fuel cell degradation models by training predictive models using historical data [146]. This journey is marked by diverse methodologies, each offering unique capabilities and insights into PEFC behavior under different operating conditions. Table 1 summarizes the existing data-driven prognostics methods for PEFC.

These models vary in the datasets from which they are built. While most of the studies use publicly available datasets such as the PHM 2014 dataset [147] and the TJU fuel cell dataset [148]; proprietary datasets might be used.

The first breakthroughs involved the implementation of an adaptive neuro fuzzy inference system (ANFIS) in learning temporal variations in stack voltage and predicting future values. Silva et al. introduced voltage data segregation into normal operation and transient perturbation segments to improve the predictive capability of ANFIS [149]. Long-term robust predictions were attained for more than 500 hours under steady-state and quasi-dynamic conditions. This work also pointed out the problem of choosing optimal parameters, usually performed in a heuristic manner. Comparative studies further explored feed-forward neural networks (FFNN) and autoregressive neural networks (ARNN), laying the groundwork for hybrid adaptive systems [150]. Further, extreme learning machines (ELMs) and deep belief networks extended the family of methods for short-term voltage prediction [151].

Wavelet-transform-based techniques ushered in a new wave of long-term predictions. Results presented by Javed et al. showed that the integration of artificial neural network (ANN) with Wavelet-transform (WT) theory, i.e., Summation-Wavelet Extreme Learning Machine (SW-ELM), provided better robustness [152]. Further improvement in mitigating uncertainties within degradation and operation conditions was introduced to SW-ELM by incorporating constraint and ensemble methods. These ensemble methods provided robust RUL predictions with error minimization, setting a benchmark for long-term performance monitoring under both steady-state and dynamic conditions.

Following the work based on WT principles, research moved towards ensemble models using the group method of data handling (GMDH) networks, enhanced ELMs, and nonlinear autoregressive exogenous model (NARX) networks [153]. So-called decomposition models decomposed complex signals into multi-timescale features and independently predicts future state-of-health metrics (for example, voltage degradation or RUL) at its own timescale, and those multi-timescale forecasts are subsequently integrated into a single, robust life-prediction output. Refinement with other optimization techniques, such as genetic algorithm (GA) and self-adaptive differential evolution, made sure to reach betterment in parameter setting and predictive performance [154]. However, physical interpretations of decomposed features are generally still required and challenging.

Toward recent developments of these data-driven models, innovation in neural network methods emerged. The capabilities of this type of methods have grown significantly with the widened power of techniques in ML methods based on recurrent neural network (RNN): long-short term memory (LSTM) and gated recurrent unit (GRU) networks, which efficiently handle sequential data through time dependencies captured in the time series [121, 155]. Building on this, other approaches such as convolutional neural networks (CNNs) have also gained attention for improving the performance of the prognostic models. Recent efforts have focused on fusing CNNs with LSTMs, utilizing transfer learning, and integrating probabilistic prognostics to model prediction uncertainty [156-159]. Meraghni et al. introduced a digital-twin (DT) approach for RUL prediction, utilizing Stacked Denoising Autoencoders (SDA). By decomposing voltage signals into level, trend, seasonal, and residual components, the DT reflected PEFC behavior [160].

Echo State Networks (ESN) introduced a paradigm shift in their reservoir architecture to decrease the computational complexity but at the same time model dynamic topology. This approach is employed in the prognostic of PEFCs by Morando et al. [161]. More in-depth development has incorporated Markov Chains and wavelet transforms discretely, extending the ability of ESN's for the prediction of degradation variation under load profiles [162].

The use of transformer models, equipped with multi-head attention mechanisms, has emerged to make predictions by modeling long-term dependencies and has shown a good performance [163]. Tian et al. employed the Informer model, a transformer-based architecture optimized for long-term sequence prediction in PEFC degradation. Using ProbSparse self-attention and distilling operations, the model effectively handles long-sequence dependencies. Tested on laboratory and real-world datasets, the Informer outperformed RNN-based models such as LSTM and GRU, achieving superior accuracy for extended prediction horizons (up to 2 and 34 hours). However, its long training times and performance degradation for very extended horizons limit real-time applicability. Additionally, reliance on preprocessed data raises concerns about adaptability to noisy, real-world conditions. Future work should address runtime efficiency and robustness in diverse operational scenarios [164]. Yang et al. conducts 1000 h dynamic durability experiments on a 5-kW stack using 16 features, then compares LSTM vs Attention-LSTM for making performance prediction. Voltage degradation speeds span 25–60 µV h⁻¹; LSTM tracks transients better, while Attention-LSTM excels for ≥200 h advanced forecasts with more stable accuracy across current modes [165]. Chou and Wang adapt the Temporal Fusion Transformer to future stack-voltage trajectories and multi-horizon RUL inference on the FC1/FC2 datasets. They stress two evaluation modes: varying the starting prediction point (how early you forecast) and the future horizon (how far ahead). Reported relative errors (RE) are as low as 0.0060 %–0.0148 % on FC1 (depending on training length of 300–600 h), 0.0326 % on FC2 in matched scenarios, and up to 1.65% under tougher settings-evidence that attention-based sequence models are now mainstream and competitive for RUL forecasting, especially when horizons extend far beyond the training window [166].

However, as it stands, all these approaches require large computational resources and training data sets, the latter generally involve time-consuming collection and processing. The determination of key features that detect system failures and their isolation is a major challenge. Transitions of operating states give rise to false alarms, hence decreasing the diagnostic effectiveness. Besides, generalization is very restricted for such methods, increasing time and effort in data processing and model refinement. However, the ever-improving research into advanced data-driven techniques continuously expands the capabilities of diagnostics and prognostics in PEFC systems.

Table 1. Data-driven prognostics methods.

| Model | Prediction Horizon | Load profile | HI | Ref. |
|---|---|---|---|---|
| DWT | Long-term | Static/Quasi-dynamic | Power | [114] |
| Grid-LSTM | Short-term | Dynamic | Voltage | [167] |
| GPSS | Short-term | Static | Voltage | [168] |
| ESN | Short-term/Long-term | Static/Quasi-dynamic | Voltage | [169] |
| MIMO-ESN | Short-term/Long-term | Static/Quasi-dynamic | Voltage | [121] |
| LSTM/GRU+attention | Short-term | Dynamic | Voltage | [170] |
| ESN | Short-term/Long-term | Dynamic | Relative Power-Loss Rate (RPLR) | [103] |
| DWT+EESN | Long-term | Dynamic | RPLR | [171] |
| LSTM | Short-term/Long-term | Static/Quasi-dynamic | Voltage | [172] |
| LSTM | Short-term/Long-term | Static/Quasi-dynamic | Voltage | [155] |
| LSTM+auto-encoder | Short-term | Dynamic | Voltage | [142] |
| Att-DNN | Short-term | Dynamic | Reconstructed Virtual Voltage | [173] |
| Dilated CNN with attention | Short-term/Long-term | Static/Dynamic | Voltage | [174] |
| TLTNN | Short-term | Static/Quasi-dynamic | Voltage | [175] |
| LSTM-GPR | Short-term | Static/Quasi-dynamic | Voltage | [176] |
| CNN-LSTM | Short-term | Static/Quasi-dynamic | Voltage | [177] |
| CNN, GRU, LSTM, ELM | Short-term | Dynamic | Voltage | [178] |
| TCN, LSTM, transfer learning | Short-term | Dynamic | Voltage | [158] |
| MISO-BiLSTM | Short-term | Static | Voltage | [179] |
| ESN-CRJ + AFS | Short-term/Long-term | Static/Quasi-dynamic | Voltage | [180] |
| Symbolic-based GRU | Short-term | Dynamic | Extracted HI | [8] |
| CEEMDAN + ARIMA+GRU | Short-term | Static/Quasi-dynamic | Voltage | [181] |
| LSTM | Short-term | Static/Quasi-dynamic | Calculated aging index | [182] |
| NARNN+LSTM | Short-term | Dynamic | Voltage | [183] |
| MIMO-CRJ | Short-term/Long-term | Static/Quasi-dynamic | Voltage | [184] |
| CNN-BiRNN | Long-term | Static | Voltage | [185] |
| Transformer | Long-term | Dynamic | Voltage, RVLR | [186] |
| ACO-LSTM | Short-term | Static/Quasi-dynamic | Power | [187] |
| PSO-ELM | Long-term | Dynamic | Voltage | [188] |
| XGB+RF | Short-term | Static/Quasi-dynamic | Voltage | [189] |
| LSTM; Attention-LSTM | Short-term/Mid-term | Dynamic | Voltage | [165] |
| Transformer | Short-term | Dynamic | Voltage | [190] |
| Temporal Fusion Transformer | Mid-term | Static/Quasi-dynamic | Voltage | [166] |

### 3.2.3. Hybrid Approaches

The lack of physical meaning of data-driven models has led to the development of hybrid approaches to prognostics of PEFCs. There are three categories of hybrid prognostic approach that can be distinguished depending on the combination of model-based and data-driven techniques [129, 191]. In the first category, model-based methods are utilized to extract degradation features, and then data-driven techniques carry out the degradation trend prediction with the estimation of RUL, as presented in [115, 192-194]. In the second category, data-driven approaches are directly applied to fit the degradation data or models; then model-based methods are applied for estimating the RUL [195, 196]. In the third category, several model-based and data-driven approaches are applied, whose individual outcomes are combined in a weighted manner to obtain the final prognostics [197]. A summary of existing hybrid methods is presented in Table 2. Despite the promising performance of hybrid approaches, open challenges in this field remain such as how to design the hybrid strategy, including the choice of applied physical models and data-driven models, as well as their combination.

Table 2. The list of hybrid models

| Method | Model type | Prediction Horizon | Load profile | HI | Ref. |
|---|---|---|---|---|---|
| Type 1 | PF + LSTM | Long-term | Static | Voltage | [198] |
| Type 1 | PF + LSTM | Long-term | Static | Voltage | [199] |
| Type 1 | AEKF+ NARX | Short-term/ Long-term | Static/Quasi-dynamic | Voltage | [200] |
| Type 1 | EKF + LSTM | Long-term | Static/Quasi-dynamic/ Dynamic | Voltage + $R_{cell}$ + $i_0$ | [201] |
| Type 1 | Semi-empirical +RNN | Short-term/ Long-term | Dynamic | Voltage | [202] |
| Type 1 | EECM+SVR | Short-term | Static/Quasi-dynamic | Voltage | [203] |
| Type 2 | LSSVM-RPF | Long-term | Static | Voltage | [196] |
| Type 2$^{i}$ | PINN | Long-term | Static/Quasi-dynamic | Voltage | [204] |
| Type 3 | ANFIS+AUKF | Long-term | Static/Quasi-dynamic | Voltage | [195] |
| Type 3 | SDA | Short-term/ Long-term | Static | Voltage | [160] |
| Type 3 | PA-VR + GRU | Short-term / Mid term | Static/Quasi-dynamic | Voltage | [205] |
| Type 3 | PF+ RF | Short-term / long-term | Static/Quasi-dynamic | Voltage | [206] |

$^{i}$physics-constrained learning

## 4. Current Implementation and Challenges

### 4.1. Data Structure and Metadata

Given that PHM models rely critically on high-quality data for training and validation, a robust data management framework is imperative. Adopting the FAIR principles, ensuring data is Findable, Accessible, Interoperable, and Reusable, offers a systematic route to surmount data limitations and boost model performance [207]. In PEM fuel-cell research, where experiments, simulations, and real-time monitoring produce vast, heterogeneous datasets, these principles are particularly vital. To this end, Dreger et al. [208] devised a self-describing Neo4j graph database built upon an EMMO ontology, in which every experimental entity carries a persistent identifier and is linked via provenance-capturing relationships. This unified schema makes data Findable (through global IDs and ontology typing), Accessible (via cypher queries or JSON-LD exports), Interoperable (by using shared vocabularies and standardized units), and Reusable (with comprehensive embedded metadata).

European initiatives, such as the European Open Science Cloud (EOSC), have been instrumental in establishing standards for data curation and sharing, thereby ensuring that critical information on degradation mechanisms, material properties, and operational performance is consistently documented with persistent identifiers and comprehensive metadata [209]. This harmonized data infrastructure not only enhances reproducibility and facilitates cross-disciplinary research, but it also provides a robust foundation for developing advanced prognostic models that can predict fuel cell degradation with greater accuracy.

Despite these advantages, implementing FAIR data practices comes with several challenges [210]. The vast diversity and volume of data, ranging from high-resolution imaging and electrochemical impedance measurements to dynamic operational data, are often stored in disparate repositories with inconsistent naming conventions and metadata standards, hindering seamless integration. Transitioning from a "my data" mindset to a collaborative "our data" framework demands nothing less than a cultural change, as well as extensive training paired with substantial financial and technological resources. Compared to more mature fields with established data-sharing practices, PEFC research is still developing its FAIR data infrastructure. Overcoming these hurdles is critical, as an effective FAIR data framework could significantly enhance PHM by providing interoperable, multi-source datasets that enable more accurate degradation assessments and predictive maintenance models. In turn, such data-driven approaches could substantially improve the performance and reliability of PEM fuel cell systems in real-world applications.

## 4.2. Real-Time Diagnostics and Prognostics

Real-time health monitoring of PEFCs has progressed from benchtop proof-of-concepts to fully integrated on-board sensor arrays to optimize the efficiency, reliability, and durability in operation. Examples include infrared cameras at the laboratory scale for thermal pictures of fuel cell stacks to get necessary information on temperature distribution and hot spots, possibly affecting performance and durability [167]. The on-board integrated sensors are now capable of monitoring essential parameters that inform the health assessment of the cell, which include temperature, humidity, pressure, voltage, and current density. In another work, Lee et al. developed a novel five-in-one micro-sensor system for in-situ temperature, voltage, pressure, fuel flow, and current measurement inside high temperature fuel cells (HFCs) [211]. Such multi-functional sensors grant a more holistic view of the operational state of the cell and provide valuable data for real-time adjustments and predictive maintenance. Figure 4 illustrates the first 1200 s of continuous PEFC operation recorded at 1 Hz, plotting cell voltage, cell current, anode dew-point water temperature, and anode inlet pressure in four vertically stacked panels [148]. Over this interval, the voltage oscillated between 0.564 V and 0.949 V ($\sigma$ = 0.084 V) in response to rapid load changes, while the current varied from 0 A to 35.54 A ($\sigma$ = 9.39 A). In contrast, the anode water dew-point temperature drifted slowly between 54.40 °C and 54.92 °C ($\sigma$ = 0.107 °C) and the inlet pressure fluctuated modestly from 109.09 kPa to 111.12 kPa ($\sigma$ = 0.185 kPa). These traces reveal two distinct time-scales, namely sub-second electrochemical transients and multi-second thermal and pressure drifts, which underscore the importance of multi-scale monitoring strategies capable of capturing both fast and slow dynamics for comprehensive PEFC diagnostics and operational optimization.

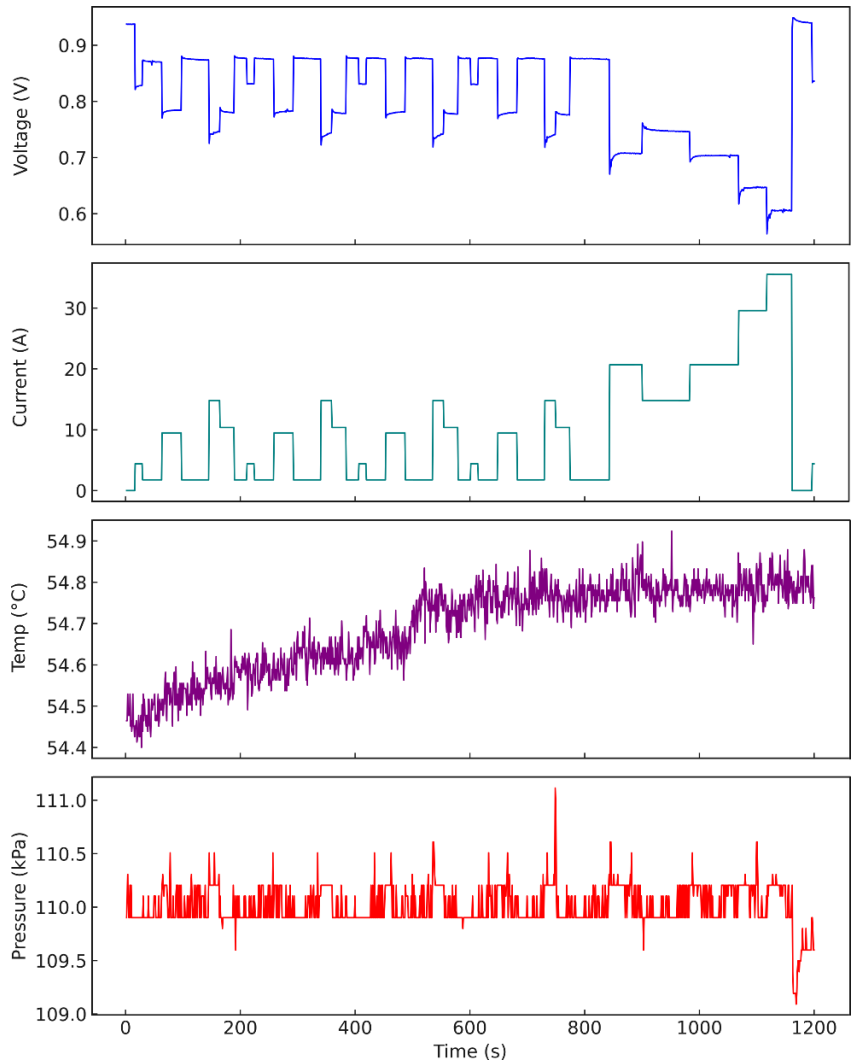

Figure 4. Time-series data from a continuous PEFC endurance test, sampled at 1 Hz over 1200 s, voltage, current, anode dew-point water temperature, and anode inlet pressure [148].

Data acquisition (DAQ) systems designed for real-time temperature monitoring, such as those proposed for air-cooled polymer electrolyte fuel cells (AC-PEFC), provide accurate control of operating temperatures and enable rapid response to thermal fluctuations. Another emerging methodology is hardware-in-the-loop (HIL) simulation testing, which addresses traditional experimental challenges by directly embedding physical components within a simulated environment. This hybrid approach thus combines the strengths of physical and digital simulation, reducing reliance on an extensive laboratory setup, while enhancing test accuracy and efficiency [212, 213]. HIL allows a smoothening of the testing cycle and the possibility of making tests more intuitive by running a real-time exchange of data between the simulated model and the actual hardware. This, in turn, provides a quicker way of prototyping and validation.

Accelerated stress testing (AST) further helps by applying controlled stresses beyond normal operation levels, hence accelerating the degradation processes and allowing the gathering of data on degradation in a short time. This is a particularly useful technology for assessing the long-term reliability of the PEFCs over much shorter test periods. For instance, Tian et al. applied AST to investigate automotive fuel cell durability and illustrated that ECSA can be used as a reliable measure of degradation for the MEA [214]. They developed a predictive model for stability and degradation rates by setting thresholds for ECSA. This way, the usefulness of AST in accelerating durability testing and ensuring the cells meet the longevity requirements in real-life operating environments, is underlined. Despite their effectiveness, there is still a pressing need for standardized and representative AST protocols created expressly for PEFC stacks. While these tests are not entirely reflective of actual vehicle operational conditions, they do provide a consistent framework for comparison. This need for commonality has resulted in the widespread use of standardized driving cycles, such as the New European Driving Cycle (NEDC) and the Worldwide Harmonized Light Vehicle Test Procedure (WLTC), to analyze metrics such as hydrogen consumption. However, uniform AST techniques for both PEFC stacks and systems have not yet been created.

The advanced sensor arrays integrated with real-time health monitoring using HIL simulation, AST, and diagnostic techniques such as EIS represent a huge step forward with PEFC technology. These innovations prolonged the operational life, economic viability, and scalability of fuel cells in applications ranging from automotive to stationary power systems. With more research in store to fine-tune all these methods and make the capability for real-time monitoring systems even more robust, PEFCs are getting ready to gain a greater share in sustainable energy infrastructures.

### 4.3. Gaps in the Action Phase

In the action phase of PHM for PEFCs, decision-making processes following prognostic analysis aim to proactively implement interventions that enhance system longevity and reliability. This phase involves several practical application scenarios, in which PHM strategies can be effectively applied. Examples of such scenarios include multi-stack management (coordinating multiple fuel cell stacks to balance load and wear), energy management systems (EMS) that optimize power distribution and operational efficiency, and predictive maintenance scheduling, which plans maintenance activities based on forecasted component health. Among these scenarios, multi-stack management, which we now define specifically for transportation PEFCs, plays a central role in balancing power output and wear across multiple, modular stacks. In transportation PEFC applications, multi-stack operation is defined as the control of several self-contained fuel-cell stacks, each a modular assembly of discrete cells, wired in parallel or series to meet dynamic drive power requirements. As envisioned in recent system-architecture studies, a multi-stack combines several discrete stack modules and auxiliaries to effect power scaling without resorting to oversized single stacks [215]. Practically, automotive prototypes such as the multi-pack system introduced by Becherif et al. [216] used parallel low-power stacks to supply peak power, while enabling module-level redundancy and easy maintenance. Control algorithms then dynamically allocate transient loads according to each stack's state-of-health, balancing degradation throughout the system and ensuring continued operation even if one stack fails to perform as anticipated, a feature essential to the durability and reliability that real-world drive cycles demand.

Building on this definition, multi-stack management aims at optimizing the operational lifespan of systems composed of multiple fuel-cell stacks [216, 217]. In such multi-stack configurations, durability is not solely dependent on the condition of an individual stack but rather on the cooperative management of all stacks to achieve balanced power output and minimize degradation. By distributing power demands across multiple stacks according to their respective health states, multi-stack management enables more efficient use of the system's resources and mitigates accelerated wear on any single stack [6, 218]. In recent studies, multi-stack management has shown promise in extending system longevity through dynamic control algorithms. For instance, optimization methods such as those employed by Chretien et al. [219] utilize mathematical models to allocate power among stacks, adjusting for each stack's degradation level and predicted RUL. This approach treats the distribution of workload as an assignment optimization problem, seeking to balance power output, while minimizing the impact of degradation on any single stack. Despite these advancements, current multi-stack control methods face notable challenges. Most approaches are developed for constant load conditions, limiting their application in variable-demand scenarios, where not all stacks need to operate continuously. In practical applications, such as electric vehicles and stationary power systems, multi-stack management strategies must dynamically respond to changing operational conditions and adjust power allocation in real time [220].

EMS represents another category with broad implications for the action phase of PHM. EMSs coordinate load-sharing between fuel cells and other energy sources, optimizing the load distribution to reduce fuel cell stress. Zuo et al. reviewed energy EMSs in the action phase of PHM for fuel cell systems, emphasizing their role in optimizing operations, extending lifespan, and reducing lifecycle costs. They classify EMS approaches based on objectives such as fuel economy, device longevity, and operational cost reduction, particularly within hybrid systems and fuel cell powertrains [221]. Although research in EMS design has explored hybrid fuel cell and battery systems, most studies, such as those by Yue et al. [222], have relied on degradation models that set rigid constraints rather than dynamically adjusting to real-time conditions.

Maintenance strategies have evolved significantly, transitioning from corrective methods to advanced prescriptive approaches, as illustrated in Figure 5. Corrective maintenance involves addressing failures after they occur, without proactive intervention, resulting in potentially costly downtime and reduced system reliability. Preventive maintenance has been emerging as a widely adopted strategy in commercial applications due to its straightforward implementation and capability to minimize unplanned downtime. This approach involves scheduling regular intervals for servicing critical components. However, neither corrective nor preventive maintenance accounts for real-time system health, limiting their effectiveness in proactively avoiding unexpected failures. Condition-based maintenance represents an advancement by utilizing real-time monitoring of the State of Health (SoH). It triggers maintenance actions based on predetermined thresholds, thereby enhancing the timely detection of issues before failures occur.

Predictive maintenance further advances maintenance strategies by estimating the RUL of components. This enables maintenance to be scheduled more precisely, thereby reducing downtime, minimizing operational disruptions, and extending asset lifespan. Techniques successfully applied in other domains can inform the development of predictive maintenance strategies tailored for PEFCs. For example, Meng et al. demonstrated, how optimized maintenance scheduling in battery systems can effectively lower operational costs [223], while Nguyen et al. proposed an LSTM-based framework for forecasting engine failures to guide replacement and inventory planning [224]. In industrial machinery, IoT-enabled predictive maintenance has been widely adopted due to its capabilities in real-time condition monitoring and intelligent diagnostics [160]. Although these approaches have proven successful, their adaptation to PEFC systems requires careful consideration of the unique degradation mechanisms, environmental conditions, and load profiles characteristic of fuel cell technologies. Despite growing interest and successful applications in sectors such as aerospace, energy storage, and micro-electro-mechanical systems (MEMS), predictive maintenance for PEFCs is still at an early stage [225, 226]. As highlighted by Dirkes et al. [6], beyond accurate scheduling, there is a need to strategically manage and potentially extend RUL to enhance long-term system performance.

The most advanced method, prescriptive maintenance, integrates maintenance actions directly with strategic decision-making processes. It employs a structured approach that utilizes real-time data analytics and prognostics to actively prescribe optimal maintenance actions, further extending system reliability and performance (as shown in Figure 5) [6].

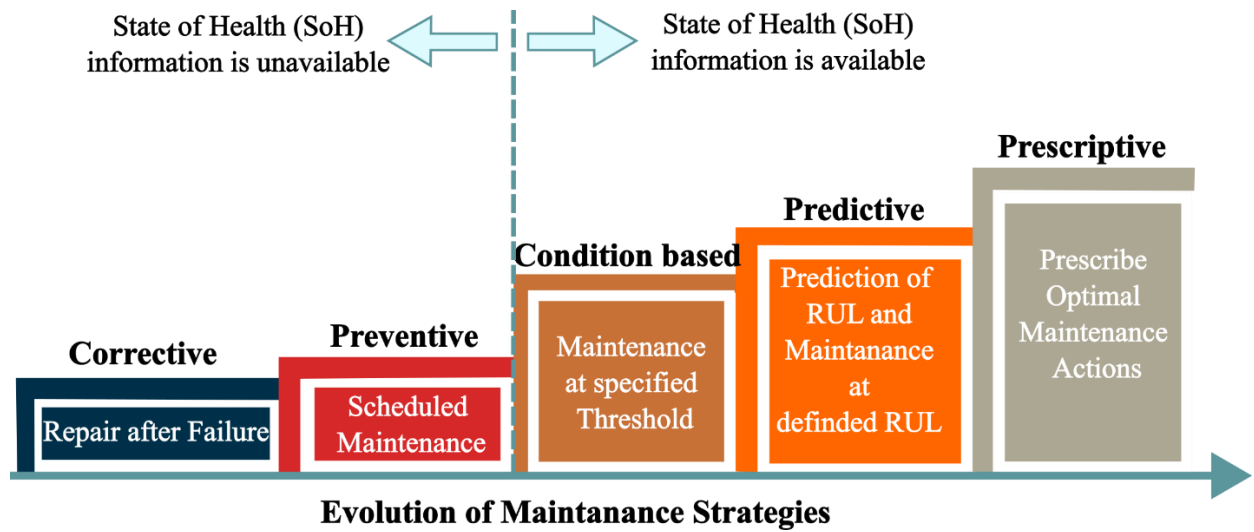

Figure 5. The schematic of maintenance strategies evolution

While substantial research efforts have targeted each category in the action phase of PHM for PEFCs, the field still lacks an integrated, data-driven framework that can synchronize these use cases effectively. Multi-stack management, energy management systems, and predictive maintenance all need to evolve beyond static models and develop adaptive algorithms that incorporate real-time prognostics. For PEFCs to achieve their full operational potential, future work needs to focus on refining degradation models, improving computational efficiency, and creating robust feedback loops between prediction and action. These advancements will facilitate a truly holistic PHM cycle capable of delivering optimized performance and extended service life across various applications.

## 5.    Key Challenges and Gaps in PHM for PEFCs

PHM for PEFCs remains a complex field with significant gaps and challenges. A fundamental issue is data scarcity, as the limited number of open-source datasets inhibit the development of reliable, generalizable models crucial for accurate health prediction and degradation analysis. Unlike other energy systems, PEFCs exhibit unique degradation mechanisms, often influenced by a range of environmental, operational, and material-specific factors. Current data is insufficient in capturing this complexity across diverse operational states, thereby compromising model robustness and predictive accuracy. Moreover, the lack of standardized data structures and comprehensive metadata further hinders progress in this field. Robust data organization and rich metadata are essential to harmonize datasets collected under varied conditions, ensuring consistency and facilitating effective cross-study integration. Such structured data not only improves discoverability and interoperability but also enables the extraction of nuanced insights necessary for advancing PHM models. Ultimately, enhancing data structuring and metadata quality can significantly boost the predictive capability of PHM models, allowing them to more accurately reflect the multifaceted degradation phenomena observed in PEFCs under variable load and environmental conditions.

While advanced action-phase PHM methods are just emerging at the individual cell and stack level, the same gaps in real-time diagnostics and adaptive control become even more pronounced when you move up to plant-wide operations or multi-unit fleets. Ideally, PHM frameworks should bridge the gap between prognostic insights and real-time decision-making, enabling automated responses such as dynamic load adjustments or predictive maintenance scheduling to extend system lifespan. However, practical implementation is still constrained by the limited integration of real-time diagnostics with adaptive control strategies, rendering the effective operational use of RUL predictions a significant, yet unresolved, challenge.

These limitations are particularly evident in complex architectures such as multi-stack fuel cell systems, where system reliability depends not only on individual stack health but also on the coordination and balance among stacks. Managing these interdependencies requires predictive tools capable of autonomously optimizing load distribution and scheduling interventions without manual oversight, capabilities that current PHM systems largely lack.

Importantly, these challenges are not exclusive to multi-stack PEFCs but are equally relevant to broader EMS, especially in hybrid configurations involving multiple energy sources (e.g., fuel cell-battery systems). In such settings, the EMS must dynamically manage energy flow, prioritize health-aware operation of each subsystem, and make optimal trade-offs between performance, efficiency, and longevity. The complexity introduced by interdependent degradation patterns and varying response dynamics underscores the need for unified, intelligent control strategies capable of holistic system-level management across diverse architectures.

System-level diagnostics in PEFCs remain underdeveloped, particularly in capturing the hierarchical interplay of degradation mechanisms across the entire system. While substantial research has focused on individual components, such as the membrane, electrode, and gas diffusion layer, there is limited understanding of how degradation propagates through multiple levels of system integration. At the most fundamental level, interactions between components (e.g., catalyst layer and membrane) can influence local performance and initiate degradation chains. These component-level effects aggregate to influence the behavior at the cell level, where cell-to-cell interactions, such as non-uniform current or temperature distribution, further complicate performance predictions.

In PEFC stacks, the interplay among individual cells introduces additional complexity, requiring monitoring and control strategies that can detect and manage imbalances. This becomes even more critical in multi-stack configurations, where stack-to-stack coordination is necessary to ensure balanced power delivery, thermal management, and load sharing. Moreover, the interaction between stacks and Balance of Plant (BoP) components, such as compressors, humidifiers, or cooling systems, can significantly affect system health. For instance, inadequate thermal management or fluctuating air supply can exacerbate degradation in otherwise healthy stacks.

Effective PHM must therefore adopt a multi-level, system-oriented perspective, recognizing that degradation in one subsystem can propagate or amplify detrimental effects and failures in others. A narrow focus on isolated components, without accounting for these hierarchical dependencies, limits the ability to develop robust and scalable health management strategies.

A further shortcoming in PHM for PEFCs is the limited focus on actionable mitigation strategies. Although PHM systems can increasingly detect and predict faults, practical solutions for actively mitigating degradation post-detection remain sparse. For instance, methods to dynamically adjust operating conditions to slow degradation, either by adjusting humidity levels, temperature, or pressure, are underdeveloped. This gap leaves a critical void in the application of PHM, as detecting issues without the means to intervene limits the value of health management insights.

In summary, the field of PHM for PEFCs must address key limitations in data availability, adaptive model development, and action-phase integration. To fully realize the potential of PHM, future efforts should focus on the creation of open-source data, the development of system-level diagnostics, as well as dynamic and responsive action-phase methods that translate RUL predictions into immediate, life-extending interventions. Addressing these critical gaps will be essential to advancing PHM for PEFCs, enabling reliable, efficient, and long-lasting performance across a range of applications.

## 6.  Future Directions

Advancing PHM for PEFCs necessitates addressing current limitations through several strategic directions.

1. Enhancing sensor integration and incorporating fuel cell systems in Internet of Things (IoT) frameworks are critical steps toward real-time, continuous performance monitoring of PEFC. Implementing advanced sensor technologies, i.e., micro- and nano-scale sensors, enables high-resolution measurement of crucial operating parameters such as temperature, pressure, humidity, and gas concentrations. These detailed insights augment the precision of diagnostic and prognostic models. When integrated into IoT systems, these sensors facilitate seamless data acquisition, transmission, and cloud analysis, enabling the facility for instant and automated response. This integration provides predictive and prescriptive maintenance practices through anomaly detection at an early stage, on the basis of intelligent decision-making to prolong system life and enhance reliability.

2. AI and ML techniques, particularly deep learning models, offer substantial potential to improve RUL predictions and diagnostic accuracy. Data-driven approaches can identify complex patterns and correlations within large datasets, enhancing the predictive capabilities of PHM systems. However, the effectiveness of these models depends on the availability of extensive, high-quality datasets, underscoring the need for collaborative efforts to develop standardized data repositories.

3. A promising direction for PEFC health management lies in the use of digital shadow and digital twin technologies. In a digital shadow, the operational data from the physical system streams into a virtual model and provides real-time insight into the state of health of the fuel cell. On the other hand, a digital twin establishes a two-way interface in which the virtual model is not just constantly informed by the incoming data but also actively shapes or regulates the physical system by taking proactive decisions.

In PHM, digital shadows enhance situational awareness through early detection of deviations in important parameters (e.g., voltage or temperature), thereby alerting operators to imminent issues. Digital twins build on this by simulating possible fault scenarios, estimating RUL, and enabling adaptive operating tactics or automated maintenance activities. These capabilities are especially beneficial in multi-stack and hybrid PEFC-battery systems, where the coordination of multiple energy sources is needed to avoid localized failure and extend total service life.

Several enabler technologies support the effective deployment of digital shadows and twins:

- Advanced Sensor Networks: Micro- and nano-sensors (see section 6.1) supply detailed real-time data on critical operating parameters.
- Data Management and FAIR Principles: Rigorous metadata standards (see section 4.1) enable interoperability for the heterogeneous data that power digital models.
- Hybrid AI Modeling: Combining data-driven approaches (see section 6.2) with physics-based models more accurately simulates advanced PEFC degradation processes.
- Edge/Cloud Computing: Real-time data processing and acquisition at scale must be accomplished through efficient and secure computing frameworks, particularly in resource-constrained or mobile environments.

While digital twins hold great promise for closed-loop PHM and proactive maintenance, they are similarly difficult in certain respects, such as keeping up real-time performance, model fidelity, and embracing current hardware. This future work must thus be focused on lightweight high-fidelity digital twin platforms that can dynamically adapt to evolving system state, facilitating action-phase integration and autonomous control for PEFC systems.

4. Advancements in material science are important for improving the durability and performance of PEFC components. However, to be of use in enabling PHM development, the materials development process must evolve. Standardized data formats, metadata, and systematic reporting of degradation behavior are lacking in traditional research, limiting its utility for prognostic modeling. To take PHM forward, material science must yield high-quality, structured information that can reflect degradation under realistic conditions. This paradigm shift would facilitate better integration of material-level expertise in system-level models of health for the purposes of better fuel cell reliability and aging prediction.

## 10. Conclusion

Prognostics and health management (PHM) has become essential to addressing the reliability and durability challenges of PEFCs. While notable progress has been made in diagnostics and RUL estimation, significant gaps remain, particularly in the integration of prognostics with real-time operational decisions and the management of multi-stack systems. Addressing these gaps will require hybrid PHM approaches that combine improved sensor technologies, advanced AI techniques, and system-level modeling. The successful deployment of robust PHM systems for PEFCs, especially in high-demand and mission-critical applications, will depend on interdisciplinary collaboration among engineers, data scientists, and materials researchers to build a more resilient and intelligent energy infrastructure.

## List of Abbreviations

Below is a consolidated list of every abbreviation defined in this manuscript, along with its full term.

| Abbreviation | Definition |
| --- | --- |
| ACO | ant colony optimization |
| ACO-LSTM | ant colony optimization – long short-term memory |
| AEKF | adaptive extended Kalman filter |
| AFS | adaptive fuzzy sampling |
| ANFIS | adaptive neuro fuzzy inference system |
| ARIMA | autoregressive integrated moving average |
| Att-DNN | attention-based deep neural network |
| AST | accelerated stress test |
| BoP | Balance of Plant |
| CCL | cathode catalyst layer |
| CEEMDAN | complete ensemble empirical mode decomposition with adaptive noise |
| CNN | convolutional neural network |
| CRJ | cycle reservoir with jump |
| DAQ | data acquisition |
| DWT | discrete wavelet transform |
| DT | digital twin / digital shadow |
| DWT+EESN | discrete wavelet transform + ensemble echo state network |
| ECEA | electrochemical surface area |
| EEM | ensemble echo state network |
| EIS | electrochemical impedance spectroscopy |

| Abbreviation | Definition |
| --- | --- |
| ELM | extreme learning machine |
| EMS | energy management system |
| EKF | extended Kalman filter |
| EOSC | European Open Science Cloud |
| ESN | echo state network |
| ESN-CRJ | echo state network – cycle reservoir with jump |
| FFNN | feed-forward neural network |
| GMDH | group method of data handling |
| GA | genetic algorithm |
| GPSS | Gaussian process state space |
| GPR | Gaussian process regression |
| GRU | gated recurrent unit |
| HFC | high-temperature fuel cell |
| HIL | hardware-in-the-loop simulation |
| HMI | Human-Machine Interface |
| $i_0$ | exchange current |
| $R_{cell}$ | Cell resistance |
| IoT | Internet of Things |
| LSTM | long short-term memory |
| MIMO-ESN | multi-input multi-output echo state network |
| MIMO-CRJ | multi-input multi-output cycle reservoir with jump |
| MISO-BiLSTM | multi-input single-output bi-directional LSTM |
| ML | machine learning |
| MEMA | membrane electrode assembly |

| Abbreviation | Definition |
| --- | --- |
| MEA | membrane electrode assembly |
| ML | machine learning |
| NARNN | nonlinear autoregressive neural network |
| NEDC | New European Driving Cycle |
| NARX | nonlinear autoregressive with exogenous input |
| NARNN | nonlinear autoregressive neural network |
| PHEV | plug-in hybrid electric vehicle (if used) |
| PA-VR | physics-based aging model considering voltage recovery |
| PF | particle filter |
| PHM | prognostics and health management |
| PEM | polymer electrolyte membrane |
| PEFC | polymer electrolyte fuel cell |
| PF+RF | particle filter + random forest |
| PF+LSTM | particle filter + long short-term memory |
| PSO-ELM | particle swarm optimization – extreme learning machine |
| PTFE | polytetrafluoroethylene |
| RF | random forest |
| XGB | extreme gradient boosting |
| RUL | remaining useful life |
| RVLR | relative voltage loss rate |
| RPLR | relative power-loss rate |
| SDA | stacked denoising autoencoder |
| SOH | state of health |
| SW-ELM | Summation-Wavelet Extreme Learning Machine |

| Abbreviation | Definition |
|---|---|
| TCN | temporal convolutional network |
| TLNN | transfer learning with transformer neural network |
| UKF | unscented Kalman filter |
| WLTC | Worldwide Harmonized Light Vehicle Test Procedure |
| WT | wavelet transform |
| EECM | electrical equivalent circuit model |
| PINN | physics informed neural network |

**Acknowledgement:**

KM, TK and ME also acknowledge the partial financial support from the European Union's Horizon Europe Research and Innovation programme, project DECODE under Grant Agreement No. 101135537. The authors also gratefully acknowledge the financial support provided by the HITEC fellowship.

**Declaration of generative AI and AI-assisted technologies in the writing process**

During the preparation of this work the authors used ChatGPT and DeepL in order to improve language. After using this tool/service, the authors reviewed and edited the content as needed and take full responsibility for the content of the publication.